\documentclass[final,authoryear,5p]{elsarticle}

\usepackage{graphicx,bm,url,color}
\graphicspath{{./fig/}{./png/}}
\newcommand{\bra}[1]{\langle #1\rangle}

\newcommand{\ybook}[3]{ (#1). {#2} (#3).}
\newcommand{\sjour}[4]{ (#1). #4. {\em #2}, submitted, arXiv:#3.}
\newcommand{\djour}[4]{ (#1). #4. {\em #2} doi:#3.}
\newcommand{\yjourNS}[5]{ (#1). #5 {\em #2}, {\em #3}, #4.}
\newcommand{\yjourN}[5]{ (#1). #5. {\em #2}, {\em #3}, #4.}
\newcommand{\yjourS}[6]{ (#1). #6 {\em #2}, {\em #3}, #4--#5.}
\newcommand{\yjour}[6]{ (#1). #6. {\em #2}, {\em #3}, #4--#5.}
\newcommand{\yarXiv}[3]{ (#1). #3, arXiv:#2.}
\newcommand{\ymedRxiv}[4]{ (#1). #4; doi: #2. URL http://medrxiv.org/content/ #2#3.}
\newcommand{\Sec}[1]{Section~\ref{#1}}
\newcommand{\Fig}[1]{Figure~\ref{#1}}
\newcommand{\Figs}[2]{Figures~\ref{#1} and \ref{#2}}
\newcommand{\Tab}[1]{Table~\ref{#1}}

\newcommand{\dd}{{\rm d} {}}
\newcommand{\Eq}[1]{Equation~(\ref{#1})}
\newcommand{\Eqs}[2]{Equations~(\ref{#1}) and~(\ref{#2})}
\newcommand{\const}{{\rm const} {}}
\newcommand{\Pe}{{\rm Pe} {}}
\newcommand{\xx}{\bm{x}}
\newcommand{\km}{\,{\rm km}}
\newcommand{\dy}{\,{\rm day}}
\newcommand{\ddy}{\,{\rm days}}

\begin{document}
\begin{frontmatter}

\title{Piecewise quadratic growth during the 2019 novel coronavirus epidemic}
\author{Axel Brandenburg}

\address{
Nordita, KTH Royal Institute of Technology and Stockholm University, SE-10691 Stockholm, Sweden\\
Published in Infectious Disease Modelling {\bf 5}, 681-690 (2020), 
doi: 10.1016/j.idm.2020.08.014
}
\ead{brandenb@nordita.org, $ $Revision: 1.79 $ $}

\begin{abstract}
The temporal growth in the number of deaths in the COVID-19 epidemic
is subexponential.
Here we show that a piecewise quadratic law provides an excellent fit
during the thirty days after the first three fatalities on
January 20 and later since the end of March 2020.
There is also a brief intermediate period of exponential growth.
During the second quadratic growth phase, the characteristic time of the
growth is about eight times shorter than in the beginning, which can be
understood as the occurrence  of separate hotspots.
Quadratic behavior can be motivated by peripheral growth when further
spreading occurs only on the outskirts of an infected region.
We also study numerical solutions of a simple epidemic model,
where the spatial extend of the system is taken into account.
To model the delayed onset outside China together with the early one in
China within a single model with minimal assumptions, we adopt an initial
condition of several hotspots, of which one reaches saturation
much earlier than the others.
At each site, quadratic growth commences when the local number of
infections has reached a certain saturation level.
The total number of deaths does then indeed follow a piecewise
quadratic behavior.
\end{abstract}


\end{frontmatter}

\parindent=0.5 cm

\section{Introduction}
The COVID-19 pandemic has attracted significant
attention among modelers of the spreading of the disease
\citep{WZ2020,WLL2020,BKW2020,Chen2020,Liang2020,Zhou2020}.
Knowing the evolution of the numbers of cases and fatalities gives
important clues about the stage and severity of the epidemic.
On theoretical grounds, one expects the number to increase
exponentially -- at least in the beginning \citep{Bri20}.
At the same time, however, control interventions lead to subexponential
growth \citep{Fenichel+11,Chowell+16,Santermans+16,Roosa+20,Tang+20},
but this is harder to quantify and to model.

In connection with COVID-19, it was noticed early on that the increase
is close to quadratic \citep{B20,ZZ20,MB20,FF20}.
While this was always thought to be a consequence of the adopted control
interventions and confinement efforts,
it was soon realized that a quadratic growth can more directly be explained
as a consequence of what is called peripheral spreading; see the
appendix of version~2 of February~14 of \cite{B20}.

The idea of peripheral growth has caught the interest of modelers in
subsequent studies \citep{Medo,Singer,Wu,Li,BK20,TM20,RB20,Blanco}.
Such an interpretation can have far-reaching consequences, because it
implies that the spreading of the disease has effectively stopped in
the bulk of some confined population.
Further spreading is only possible on the periphery, for example through
asymptomatic individuals that escaped detection.
This inevitably led to further spreading outside China through the rest
of the world.

The purpose of the present paper is to substantiate the idea of peripheral
growth through standard epidemiological modeling.
The simplest of such models is that of \cite{KMK27}.
It is now commonly referred to as the $SIR$ model, where
$S$ stands for the number of susceptible individuals,
$I$ for the number of infectious individuals, and
$R$ for the number of recovered, deceased, or immune individuals.
The spatial dimension is added to the problem by introducing a diffusion
operator \citep{Nob74,KAM85,Mur86}; see also the text book by \cite{Mur03}
for a detailed account on biological modeling in space and time.
We show that this model can explain the piecewise quadratic growth
observed during COVID-19 outbreak.
The shorter time constant during the second quadratic growth phase is
modeled as an increase in the number of separated hotspots, from which
peripheral growth occurs.
We begin, however, with a detailed discussion of the evidence for
quadratic growth during various stages of COVID-19.

\begin{figure}[t!]\begin{center}
\includegraphics[width=\columnwidth]{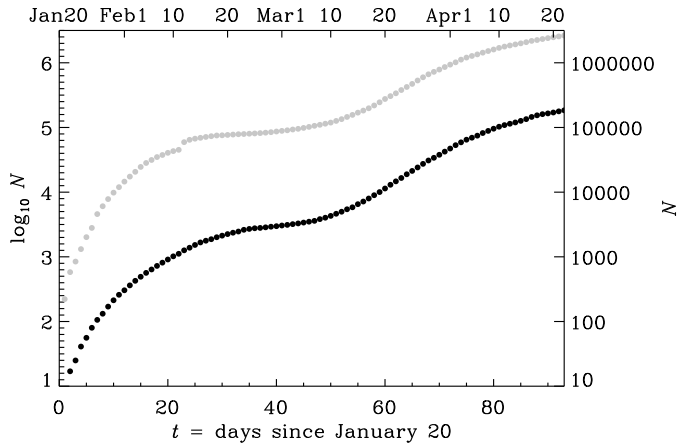}
\end{center}\caption[]{
Logarithmic representation (base 10) of the number of deaths (black
symbols) and infections (gray symbols) as a function of time since
January 20, 2020.
The actual date is given on the upper axis and the actual values
of $N$ are given on the right-hand axis.
}\label{p4}\end{figure}

\section{Quadratic versus exponential growth}
\label{Empirical}

Our primary interest is in the number of infections, but this number
is uncertain because it depends on the amount of tests that are done in
each country.
A more robust proxy is the number of deaths.
For the data after January 22, 2020, we use the worldometers
website\footnote{\url{http://www.worldometers.info/coronavirus/}},
while the data of earlier days can be found on the DEVEX
website\footnote{\url{http://www.devex.com/news/2019-ncov-outbreak-a-timeline-96396}}.
In \Fig{p4}, we show the number of infections and the number of deaths
in a semi-logarithmic representation.
Following \cite{B20}, time is here counted as the number of days
after January 20, which he identified as the date when quadratic
growth commenced.
Specifically, he found
\begin{equation}
N_{\rm fit}(t)=[(t-\mbox{Jan 20})/0.7\,\mbox{days}]^2
\;\;\;\mbox{(quadratic fit)}.\;\;
\label{death}
\end{equation}
This means that every 0.7~days, the square root of $N$ changes by one.
Such an increase is much slower than an exponential one, where instead
the logarithm of $N$ changes by one during one characteristic time.
To illuminate the quadratic growth in more detail, we consider an
example for January~30.
In that case, \Eq{death} predicted $N_{\rm fit}=(10/0.7)^2=204$, so
0.7~days later, $N_{\rm fit}^{1/2}$ changed by one (from 14.3 to 15.3)
and therefore $N_{\rm fit}=(10.7/0.7)^2=234$, corresponding to an
increase of $N$ by 30.
On February 20, i.e., 31 days after January 20, the formula predicted
$N_{\rm fit}=(31/0.7)^2=1960$, so 0.7~days later,
$N_{\rm fit}=(31.7/0.7)^2=2050$ has increased by 90.
This gives us a sense of the way $N$ increased.
These numbers agree quite well with the actual ones.

\begin{figure}[t!]\begin{center}
\includegraphics[width=\columnwidth]{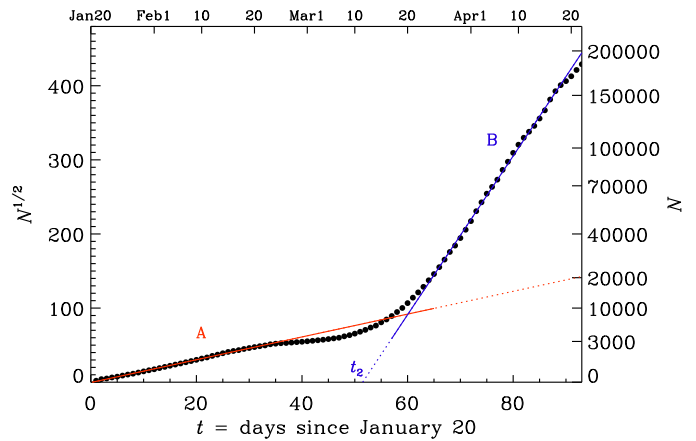}
\end{center}\caption[]{
Same as \Fig{p4}, but for the square root of $N$ (black dots).
The red line corresponds to the fit given by \Eq{death}, and the
blue line is a similar fit with parameters given in \Tab{TAB2}.
}\label{p3}\end{figure}

Real data never agree perfectly with any particular mathematical
growth law.
It is therefore important to quantify the accuracy of any such a
description.
To assess the accuracy of a description in terms of a quadratic growth
law, it is useful to plot the square root of the number of death,
$N^{1/2}$, because this quantity then increases linearly in such a
representation; see \Fig{p3}.
We immediately notice the appearance of two subranges, A and B, with
approximately quadratic growth, but different slopes.

\begin{table}[b!]\caption{
Parameters of three exponential fits.
Time is in days, starting on January 20, 2020.
}\vspace{12pt}\centerline{\begin{tabular}{crrrcc}
Interval & $t_1$ & $t_2$ & $t_0$ & $\tau$ & $\sigma$ \\
\hline
  I &  1 &  8 &$ -6.4$&  2.9  & 0.079 \\
 II & 10 & 25 &$-32.7$&  7.8  & 0.060 \\
III & 54 & 75 &$-19.6$&  8.5  & 0.021 \\
\label{TAB}\end{tabular}}\end{table}

\begin{table}[b!]\caption{
Like \Tab{TAB}, but for the parameters of the square root fits.
}\vspace{12pt}\centerline{\begin{tabular}{crrrcc}
Interval & $t_1$ & $t_2$ & $t_0$ & $\tau$ & $\sigma$ \\
\hline
A &  5 & 36 &$  0.3$&  0.65 & 0.042 \\
B & 65 & 90 &$ 51.5$&  0.093 & 0.019 \\
\label{TAB2}\end{tabular}}\end{table}

\begin{figure}[t!]\begin{center}
\includegraphics[width=\columnwidth]{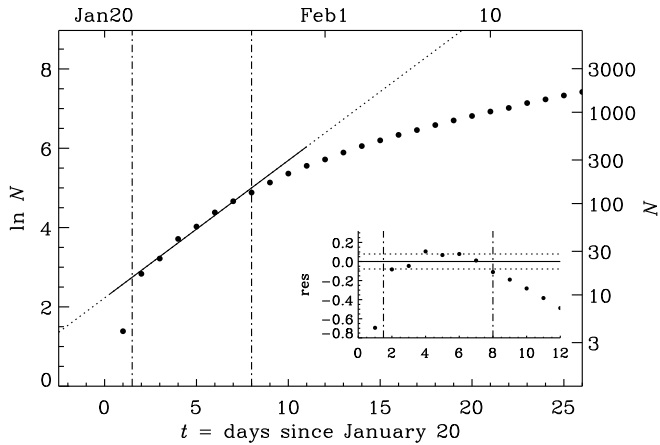}
\end{center}\caption[]{
Semi-logarithmic representation of $N$ during the early phase.
The two vertical dash-dotted line indicate the fit range where
$\ln N$ increases approximately linearly with $t$.
The inset shows the residual with the two horizontal dotted lines
indicating the value of $\pm\sigma$.
Here and in the following plots, the natural logarithm is used.
}\label{pfit_log_early}\end{figure}

Before substantiating the reality of quadratic growth, we first examine
the possibility of exponential growth during early times.
In \Fig{pfit_log_early}, we show a semi-logarithmic representation for
the end of January 2020.
We see that, even at early times, there is no convincing evidence
for exponential growth, although it is always possible to identify
an approximately constant slope during short time intervals.
The fit shown in \Fig{pfit_log_early} for January 22--28 is given by
\begin{equation}
N_{\rm fit}(t)=\exp[(t-t_0)/\tau]\quad\mbox{(exponential fit)},
\label{ExpFit}
\end{equation}
where $\tau$ is the $e$-folding time and $t_0$ is some reference time
(where the fit intersects the abscissa); see \Tab{TAB} for a summary of
the parameters.
The $e$-folding time increases from about 3 days during Interval~I
to about 8 days during Intervals~II and III; see the values of $\tau$
in \Tab{TAB}.

\begin{figure}[t!]\begin{center}
\includegraphics[width=\columnwidth]{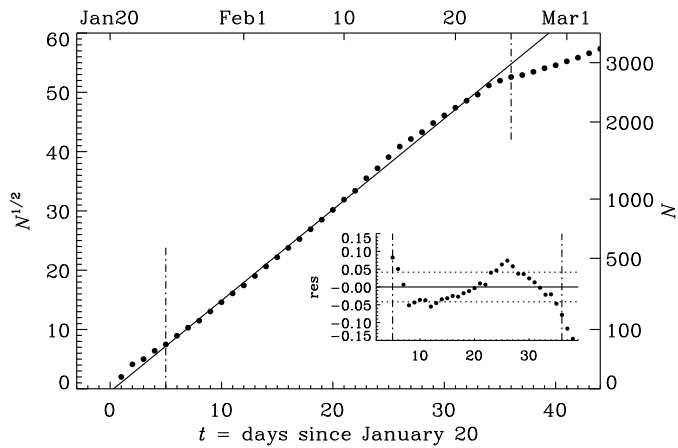}
\end{center}\caption[]{
Similar to \Fig{pfit_log_early}, but for a square root representation
of $N$ during the early phase.
}\label{pfit_sqr_early}\end{figure}

Next, to quantify the accuracy of the fits given by \Eq{ExpFit}
for limited time intervals, $t_1\leq t\leq t_2$,
we compute the relative residual as
\begin{equation}
\mbox{res}=[N(t)/N_{\rm fit}(t)-1].
\label{Res}
\end{equation}
This residual is shown as an inset to \Fig{pfit_log_early}, and its
standard deviation, $\sigma=\bra{\left(\mbox{res}\right)^2}^{1/2}$,
is indicated by dotted lines.
Here, angle brackets denote averaging over the time span
$t_1\leq t\leq t_2$.
For Interval~I, $\sigma$ is approximately 8\%; see \Tab{TAB}.
For the quadratic fit, we write
\begin{equation}
N_{\rm fit}(t)=[(t-t_0)/\tau]^2\quad\mbox{(quadratic fit)},
\label{QuadraticFit}
\end{equation}
where $\tau$ is again a characteristic time and $t_0$ is a reference time,
where the fit intersects the abscissa.
This is fairly analogous to the exponential fit, but the meaning of $\tau$
is different; see \Tab{TAB2} for the parameters of the quadratic fits.
The residual of \Eq{Res} applies to both types of fits.
The quadratic fit of \Eq{death} has a standard deviation of only 4\%
over a much longer time interval of about 30 days in Interval~A (see
\Fig{pfit_sqr_early}) compared to the exponential fits for Interval~I
(\Fig{pfit_log_early}) and II (see \Tab{TAB} for the parameters).

\begin{figure}[t!]\begin{center}
\includegraphics[width=\columnwidth]{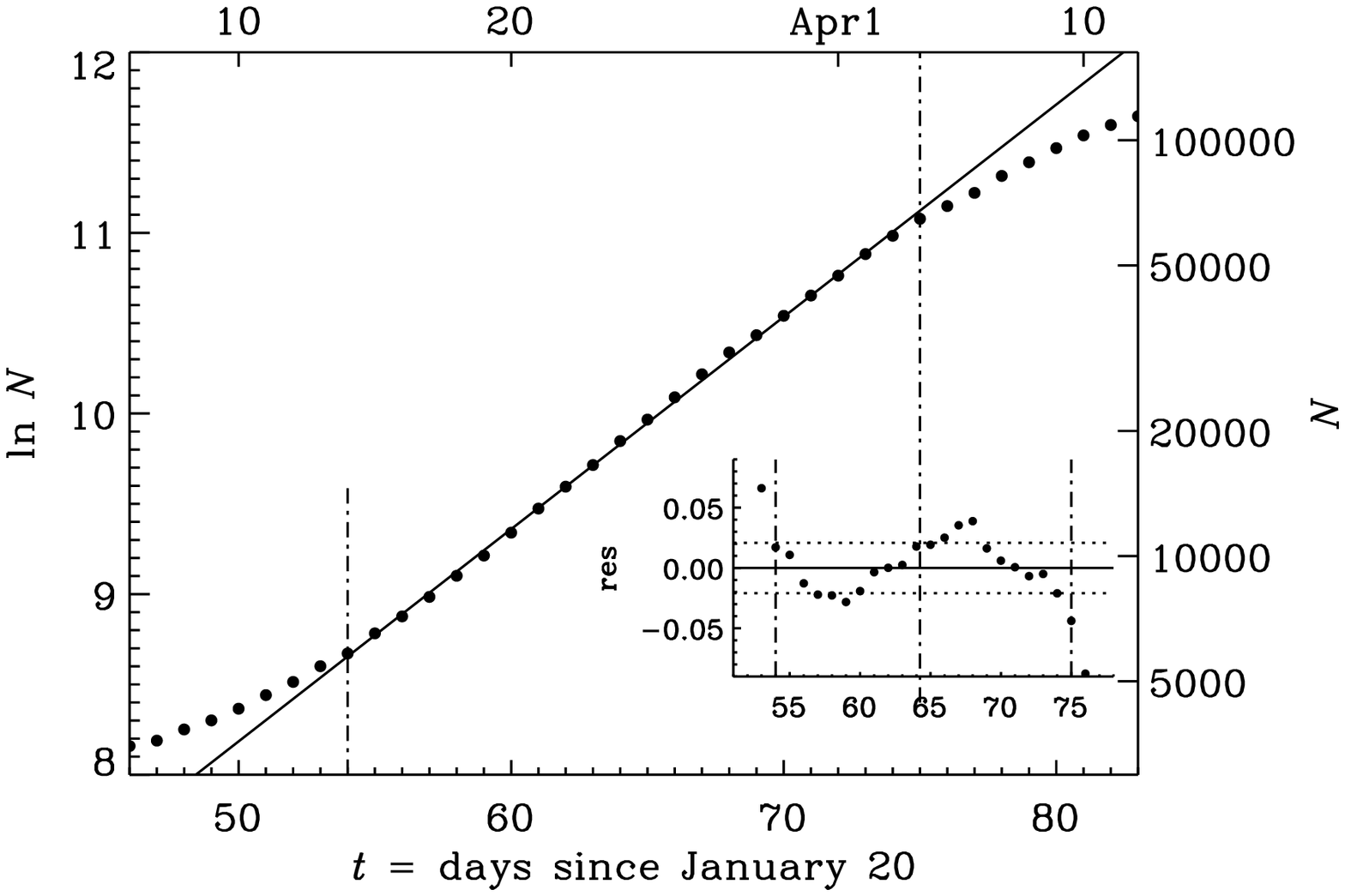}
\end{center}\caption[]{
Similar to \Fig{pfit_log_early}, but for the late phase (Interval~III).
}\label{pfit_log_late}\end{figure}

\begin{figure}[t!]\begin{center}
\includegraphics[width=\columnwidth]{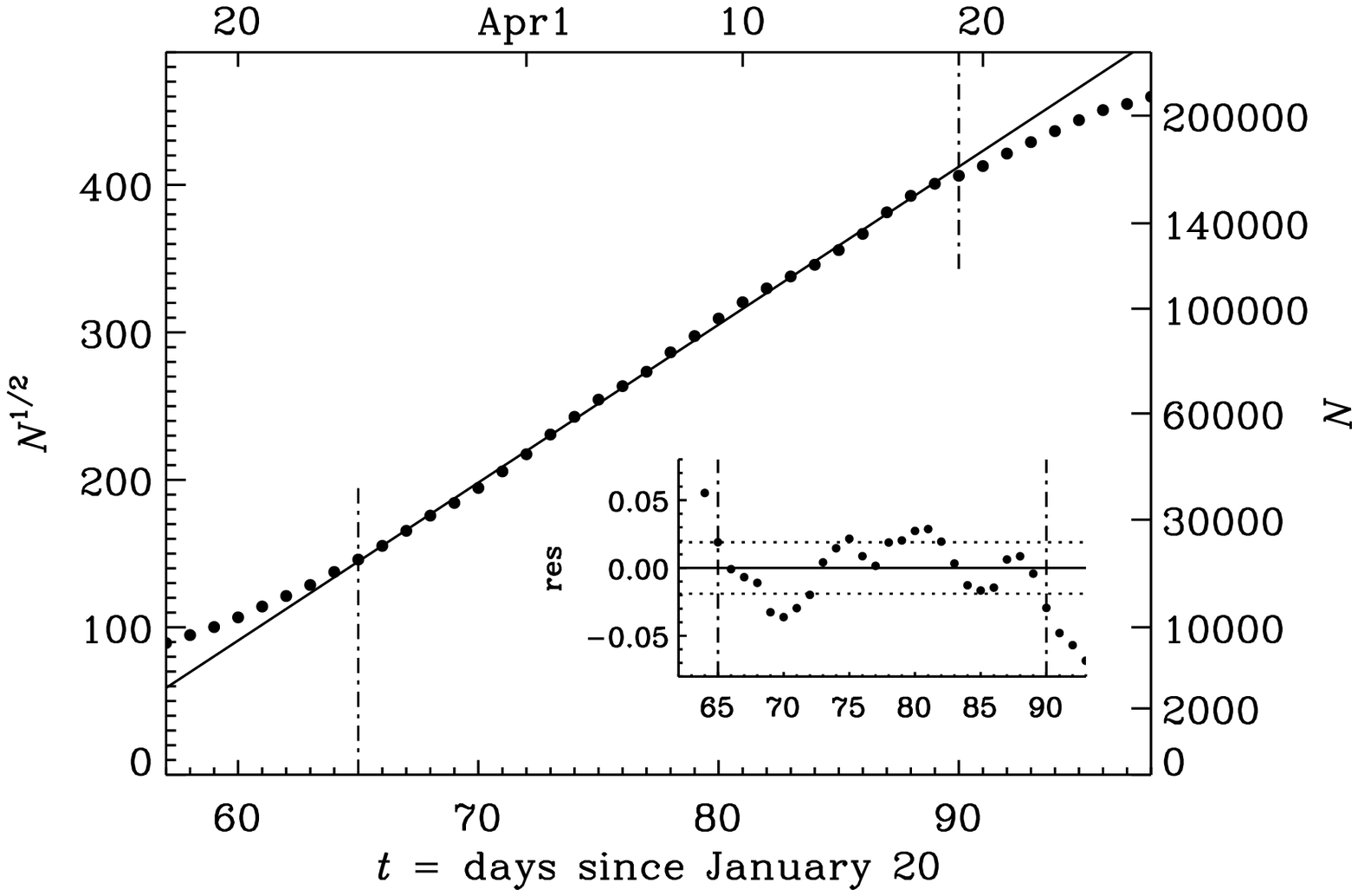}
\end{center}\caption[]{
Similar to \Fig{pfit_sqr_early}, but for the late phase (Interval~B).
}\label{pfit_sqr_late}\end{figure}

As mentioned before, there is an intermediate phase (Interval~III),
where the growth is indeed approximately exponential with a value
of $\sigma$ of about 2\% for about 20 days; see \Fig{pfit_log_late}.
This stage is followed by a quadratic growth (Interval~B)
until the present time with $\sigma=1.9\%$.
In \Fig{pfit_sqr_late} we show such a representation.
The characteristic time is now about 0.09 days, which is eight
times shorter than the time during the early stage in Interval~A.
This means that $N^{1/2}$ changes by one every 0.09 days or by
about ten every day.
We have to remember that $N$ is now larger than for Interval~A: when
$N^{1/2}=400$ (late times in \Fig{pfit_sqr_late}), a change of $N^{1/2}$
by ten per day corresponds to a change of $N$ from $400^2=160000$ to
$410^2=168100$, i.e., a change by about $8000$ in one day.

\section{Heuristic model of peripheral spreading}
\label{Heuristic}

Contrary to the usual exponential growth, a quadratic growth can be the
result of control interventions.
It can be explained by spreading on the periphery of a bulk structure,
which can be of geometrical or sociological nature \citep{Bri19}.
In the bulk, no further infections are possible.
At the level of an epidemic model, it would correspond to a situation
where the local population density has effectively reached saturation
levels in the number of infections \citep{Bri20b}.
In the case of COVID-19, however, it is more realistic to describe this
as a state of partial confinement and isolation of individuals
or small groups of people.

\begin{figure}[t!]\begin{center}
\includegraphics[width=\columnwidth]{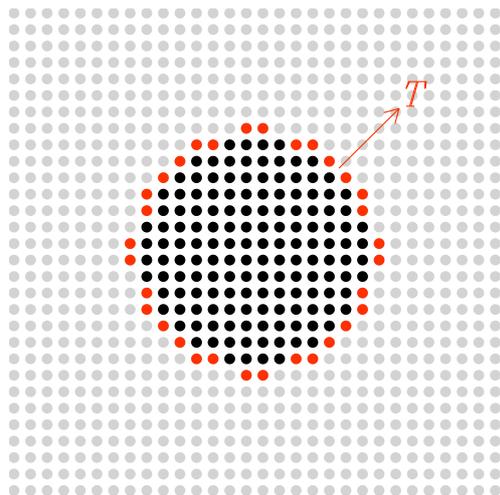}
\end{center}\caption[]{
Sketch illustrating the group of infected people confined to the bulk
(here for $N=148$, black filled symbols) and a group of people at the
periphery (here $n=32\approx3\sqrt{N}$, red filled symbols) responsible
for spreading the disease with a characteristic time scale $T$.
}\label{psketch}\end{figure}

Here we propose a model where the continued confinement efforts prevent
spreading within the bulk of the infected population, but these efforts
cannot prevent spreading on the periphery; see \Fig{psketch} for a sketch.
The rate $\dd N/\dd t$, with which $N$ increases in time, is therefore
equal to the number of infected people on the periphery divided by a
characteristic spreading time $T$.
We therefore arrive at the following simple differential equation
\begin{equation}
\frac{\dd N}{\dd t}=\frac{n}{T},
\label{dNdt}
\end{equation}
where $n$ is the number of people in a narrow strip around the periphery.
Its size scales with the ratio of the circumference ($=2\pi r$ for a
circle of radius $r$) to the square root of the area ($=\pi r^2$),
so $n\approx2\sqrt{\pi N}$.
Here, the prefactor depends on the geometry and we would have $n=4\sqrt{N}$
for a rectangular geometry.
We may therefore set $n=\alpha\sqrt{N}$, where $\alpha\approx3.5$ for
a circular geometry.
Inserting this into \Eq{dNdt} yields
\begin{equation}
\frac{\dd N}{\dd t}=\frac{\alpha}{T}\sqrt{N},
\end{equation}
with the solution
\begin{equation}
N(t)=(\alpha t/2T)^2.
\label{Nt}
\end{equation}

The empirical analysis of \Sec{Empirical} suggested a growth of the form
$N(t)=(t/\tau)^2$, with $\tau\approx0.7\,\mbox{days}$ (or about 17 hours)
and $t$ being the time in days after January 20, 2020 for Interval~A and
$\tau\approx0.093\,\mbox{days}$ for Interval~B.
For a circular geometry, this implies that the spreading times are
$T=\alpha\tau/2=1.2\,\mbox{days}$ and $0.16\,\mbox{days}$ for
Intervals~A and B, respectively.

The idea of a geometrically confined bulk with a surrounding periphery
may need to be generalized to sociological or network structures that
can follow similar patterns \citep{Ian17,Bri19,Prasse}.
In the present work, we do not make any attempts to analyze this aspect
further, but refer instead to recent work of \cite{Sanche}, who analyzed
the spatial patterns of COVID-19 during the early phase, and to the
work of \cite{ZZ20}, who also discussed spreading on a fractal network.

The increase in the slope of $N^{1/2}$ versus $t$ corresponds to a
decrease of $\tau$ by a factor of about eight for Interval~B.
This can be the result of multiple separate hotspots, each of which
display peripheral growth.
This idea will be substantiated further in the following, where we use
the spatio-temporal epidemiological model of \cite{Nob74}.
Such a model leads to radial expansion waves that propagate at constant speed.
It is therefore expected to lead to situations similar to what we have
discussed above.

\begin{figure*}[t!]\begin{center}
\includegraphics[width=\textwidth]{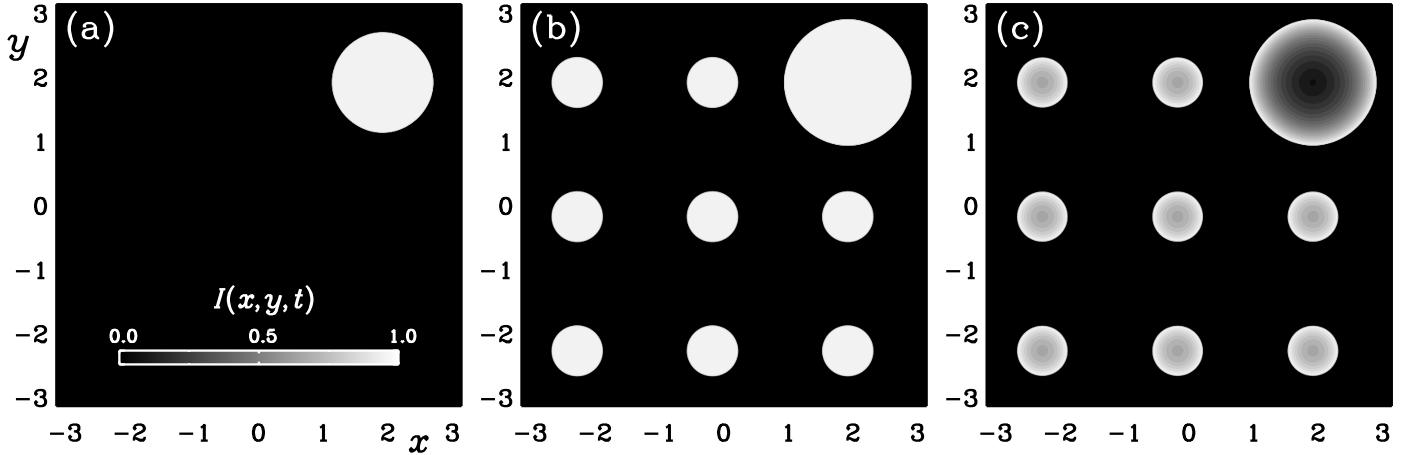}
\end{center}\caption[]{
$I(x,y,t)$ for $t=400$ and $\mu=0$ (a), $t=500$ and $\mu=0$ (a),
and $t=500$ and $\mu=5\times10^{-3}$ (c).
}\label{psav}\end{figure*}

At later times, several hotspots can merge.
This leads to a decrease in the slope, which has in fact also been observed
since the middle of May.
Corresponding plots are presented along with the datasets for the present
paper \citep{B20data}.
The discussion of such models will be postponed to a subsequent paper.

\section{Epidemiological model with spatial extent}

The $SIR$ model of \cite{KMK27} is a predator--prey type model, where the
number of prey corresponds to the susceptible individuals $S$, and the
number of predators corresponds to the infected population $I$.
The latter also spreads to their neighbors by diffusion, described by
a diffusion term $\kappa\nabla^2 I$ with $\kappa$ being a diffusion
coefficient \citep{Nob74,KAM85,Mur86}.
Finally, there is the number of diseased or recovered individuals $R$.
The quantity $I$ can be identified with the variable $N$ used in
\Sec{Empirical}.
Thus, we have
\begin{equation}
{\partial S\over\partial t}=-\lambda SI,
\label{dSdt}
\end{equation}
\begin{equation}
{\partial I\over\partial t}=\lambda SI-\mu I+\kappa\nabla^2 I,
\label{dIdt}
\end{equation}
\begin{equation}
{\partial R\over\partial t}=\mu I.
\end{equation}
In a closed domain, the total number of individuals is constant,
so $\bra{S+I+R}=\const$.
We therefore only need to solve \Eqs{dSdt}{dIdt}.

To solve \Eqs{dSdt}{dIdt}, we employ the 
{\sc Pencil Code}\footnote{\url{http://github.com/pencil-code},
doi:10.5281/zenodo.2315093}, a publicly available time stepping code
for solving partial differential equations on massively parallel architectures
\citep{PC}.
Spatial derivatives are computed from a sixth-order finite difference
formula and the third order Runge--Kutta time stepping scheme of
\cite{Wil80} is employed.
We use $4096^2$ mesh points and run the model for about 1200 time
units, which takes about six minutes with 1024 processors on a Cray XC40.
We fix the time step to be 0.05 and have checked that the solution did
not change when the time step is decreased further.
The $SIR$ model is implemented in the current version, and also
the relevant input parameter files are publicly available
\citep{B20data}.

We solve \Eqs{dSdt}{dIdt} in a two-dimensional Cartesian domain with
coordinates $\xx=(x,y)$ and periodic boundary conditions.
We characterize the domain size $L$ by the smallest wavenumber $k=2\pi/L$
that fits into the domain.

The model has three parameters: the reproduction rate $\lambda$, the rate
of recovery $\mu$, and the diffusion constant $\kappa$.
In this model, a certain fraction of $R$ could be interpreted as the number
of deaths, but this distinction will not be made in the present work.
In addition to the three aforementioned parameters, we have the spatial
and temporal coordinates, $\xx$ and $t$.
It is convenient to define nondimensional space and time coordinates
as $\tilde{\xx}=k\xx$ and $\tilde{t}=\lambda t$.
This leaves $\tilde{\mu}=\mu/\lambda$ and $\tilde{\kappa}=\kappa k^2/\lambda$
as the only nondimensional input parameters that we shall vary.
The population number is normalized by the initial number of susceptible
individuals, $S_0$, so we can define $\tilde{S}=S/S_0$, $\tilde{I}=I/S_0$,
and $\tilde{R}=R/S_0$ as the fractional (nondimensional) population
densities.
We then have $\bra{\tilde{S}+\tilde{I}+\tilde{R}}=1$ at all times.

The tildes will from now on be dropped.
In practice, this means that we always keep $\lambda=1$ and adopt for the
domain size $L=2\pi$, so $k=1$.

As initial condition, we assume $S=1$ and $I=0$, except for nine
mesh points, where we initialize $I=I_1$ on one isolated mesh point
and $I=I_2$ on eight others.
We refer to them as ``hotspot''.
We always use $I_1=10^{-6}$ for the main hotspot, which we place
at $x\approx y\approx 2$; see \Fig{psav}.
Since the growth rate is normalized
to unity, one would expect $I_1$ to reach unity in a time
$t=-\ln10^{-6}=6\times\ln10\approx14$ in the absence of saturation.
We perform different experiments using for the secondary multiple
hotspots the values $I_2=10^{-60}$, $10^{-180}$, and $10^{-300}$,
which, in the absence of saturation, would reach saturation at times
$t=60\times\ln10\approx140$, $180\times\ln10\approx410$, and
$300\times\ln10\approx700$.

We begin by studying models with $\mu=0$, but later we also consider
small nonvanishing values of $\mu$.
For most of our models, we use $\kappa=10^{-6}$, which is close to
the smallest value that is allowed at our resolution of $4096^2$.
It implies that reaction fronts are sufficiently well resolved.
Their size and speed depend on the values of $\lambda$ and $\kappa$
and can be obtained on dimensional grounds.
It is therefore useful to restore the symbol $\lambda$, even though we
have already put it to unity.
In our case, the width of the front is $\sqrt{\kappa/\lambda}=10^{-3}$.
Such an expression is typical of reaction--diffusion equations
\citep{Fis37,Kol37}.
The front speed is $c=2\sqrt{\lambda\kappa}$ \citep{Mur03}, which is
characterized by the nondimensional quantity $\Pe=c/\kappa k=2000$,
which is also known as the P\'ecl\'et number.

\begin{figure}[t!]\begin{center}
\includegraphics[width=\columnwidth]{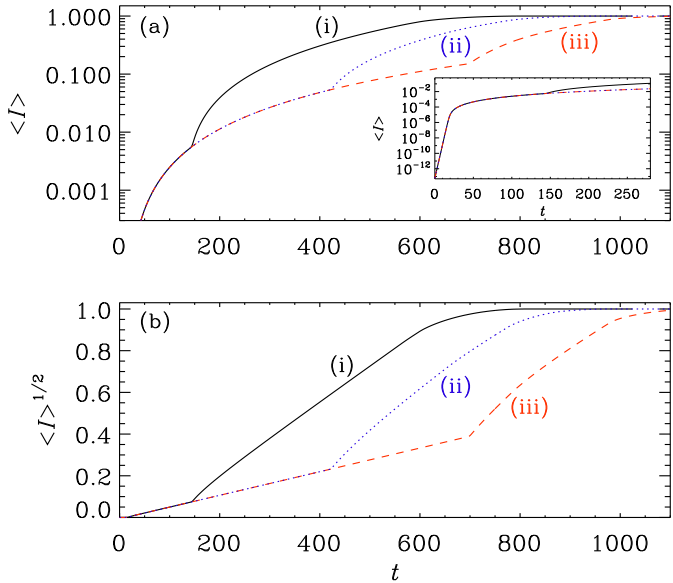}
\end{center}\caption[]{
(a) semi-logarithmic and (b) square root representations of $\bra{I}$ for
initial values (i) $I_2=10^{-60}$, (ii) $10^{-180}$, and (iii) $10^{-300}$.
}\label{pcomp}\end{figure}

\begin{figure}[t!]\begin{center}
\includegraphics[width=\columnwidth]{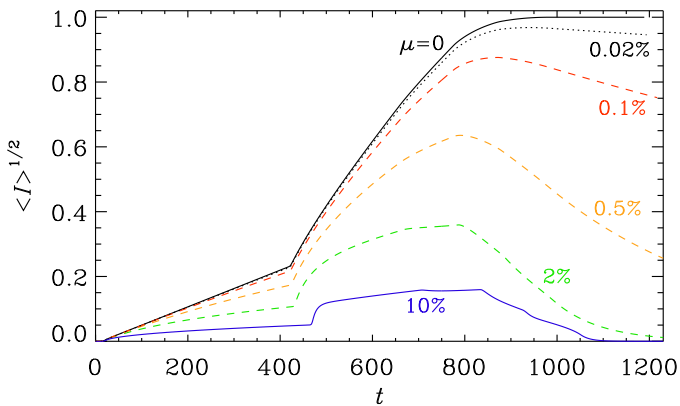}
\end{center}\caption[]{
Simulations for $\mu=0$ (black), $\mu=2\times10^{-4}$ (dotted),
$\mu=10^{-3}$ (red), $\mu=5\times10^{-3}$ (orange),
$\mu=2\times10^{-2}$ (green), and $\mu=10^{-1}$ (blue).
}\label{pcompk}\end{figure}

\begin{figure}[t!]\begin{center}
\includegraphics[width=\columnwidth]{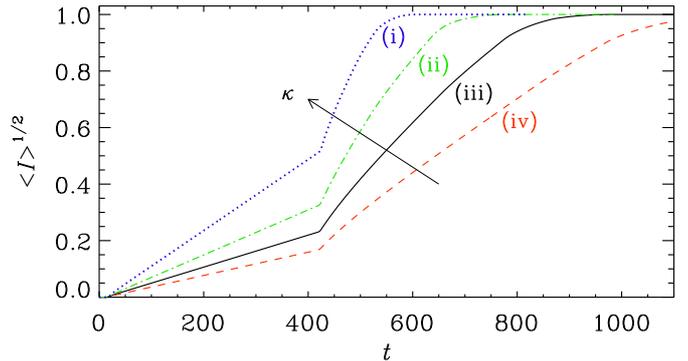}
\end{center}\caption[]{
Simulation for $\kappa=5\times10^{-6}$ (i), $\kappa=2\times10^{-6}$ (ii),
$\kappa=10^{-6}$ (iii), and $\kappa=5\times10^{-7}$ (iv).
The arrow points in the direction of increasing values of $\kappa$.
}\label{pcomp_kappa}\end{figure}

In \Fig{psav} we show gray scale visualizations of $I(x,y,t)$ at two
instants shortly before ($t=400$) and shortly after ($t=500$) the time
when the secondary hotspots become significant (shown here for $\mu=0$).
For $t=500$, we also present a case with $\mu=0.005=0.5\%$.
We see that this small value of $\mu$ hardly affects the spatial
spreading of the disease.
In the interior of the affected region, however, there is a certain
recovery, so $I(x,y,t)$ decreases again at the center of each hotspot.
We have not plotted $S(x,y,t)$, but we note that it has an essentially
complementary structure in that its value drops locally approximately by
the same amount that $I(x,y,t)$ increases.

In \Fig{pcomp}, we show the evolution of $\bra{I}$ for our three values
of $I_2$ ($=10^{-60}$, $10^{-180}$, and $10^{-300}$).
We see that at very early times, $\bra{I}$ increases exponentially.
This is seen in the inset of \Fig{pcomp}(a).
Locally, the big hotspot has reached saturation within a tiny spot,
which then begins to expand.
We recall that the primary spot had an initial value of $10^{-6}$, 
but in the inset we plot the averaged value $\bra{I}$, which can be
$4096^2\approx2\times10^{-7}$ times smaller.

At some point, the values of $I(x,y,t)$ at the secondary hotspots begin
to become significant and, because of their larger number, begin to
dominate the growth of $\bra{I}$.
This is similar to the late phase in Interval~B shown in \Fig{p3}.
By the argument presented in \Sec{Heuristic}, the slope of the graph of
$N^{1/2}$, which is proportional to $\bra{I}^{1/2}$, should scale with the
ratio of their total circumference to the square root of the total area.
Therefore, the slope should scale with the square root of the number of
secondary hotspots.
In this connection, we note that a dependence of the total reaction speed
on the number of topologically disconnected regions is also typical
of other reaction--diffusion equations and has been seen before; see
Figure~4 of \cite{BM04}.

Next, we study the effects of changing of $\mu$.
We see that already rather small values of around $\mu=2\times10^{-4}=0.02\%$
have a noticeable effect at late times, so that $\bra{I}$ reaches a maximum
as a function of time at around $t=800$.
The position of this maximum depends only weakly on the value of $\mu$.

Finally, we study the effects of changing the diffusivity $\kappa$.
The result is shown in \Fig{pcomp_kappa}.
We see that the speed of spreading, which is roughly $2\sqrt{\lambda\kappa}$,
increases with increasing diffusivity.

Having now gained some experience with this model, we can ask what would be
realistic parameters related to COVID-19.
Given that the slope of $N^{1/2}(t)$ depends on the value of $\kappa$,
one might be able to give some estimates.
Using the slopes seen in \Fig{pcomp_kappa}, we find that
$\tau^{-1}=\beta\sqrt{\lambda\kappa k^2}$, where $\beta_{\rm A}\approx0.56$
for Interval~A and $\beta_{\rm B}\approx2.8$ for Interval~B, with the
subscript denoting the interval.
Again, we have restored here the symbol $\lambda$, even though $\lambda=1$
was assumed in all of our simulations.
The ratio of the two coefficients is around five, which is nearly twice
as much as the square root of the number of secondary hotspots, which
was expected based on the heuristic argument presented in \Sec{Heuristic}.
To estimate the effective value of $\kappa$, the largest uncertainty
comes from the value of $k$, which is the inverse domain size which,
in turn, is ultimately related to the size of the affected continents
on the Earth.
Assuming that $\kappa\approx(\lambda\tau^2 k^2)^{-1}$, we see that with
$k\approx(1000\km)^{-1}$, $\lambda=(10\ddy)^{-1}$, and $\tau=1\dy$,
we have $\kappa\approx10^9\km^2/\dy$, which is much larger than the
diffusion coefficient estimated for the spreading of the Black Death in
1347, for which a diffusion coefficient of the order of $10^2\km^2/\dy$
has been estimated \citep{Nob74}.

\section{Discussion}

The present work has demonstrated that for the COVID-19 epidemic,
the available data are accurate enough to distinguish between an
early exponential growth, as was found for the 2009 A/H1N1 influenza
pandemic in Mexico City \citep{Chowell+16} and the quadratic growth
found here.
It was expected that the growth would not continue to be exponential,
and that it would gradually level off in response to changes in the
population behavior and interventions \citep{Fenichel+11}, as found
in the 2014-15 Ebola epidemic in West Africa \citep{Santermans+16}.
However, that the growth turns out to be quadratic to high accuracy
already since January~20 is rather surprising and has not previously
been predicted by any of the recently developed models of the COVID-19
epidemic \citep{Chen2020,Liang2020,Zhou2020}.
It is therefore important to verify the credibility of the officially
released data; see \cite{Rob19} for similar concerns in another context.

In the case of COVID-19, it is remarkable that the fit isolates
January~20 as a crucial date in the development of the outbreak.
At that time, the actual death toll was just three and the number of
confirmed infections just a little over 200.

It is rare that an epidemic provides us with data having such systematic
trends as in the present case.
One might wonder why this quadratic growth is not generally discussed
in the literature.
There is a large variety of theoretical models; see \cite{Chowell+16}
for a review.
Several such models have already been adapted to the COVID-19 epidemic
\citep{Chen2020,Liang2020,Zhou2020,Bri20}.
Furthermore, the idea of control interventions has been discussed in
detail \citep{Fenichel+11}, but the present epidemic provides us with
an unprecedentedly rich data record with large numbers of infections
occurring on an extremely short timescale.
All this contributes to having made the quadratic growth so apparent.

By the time the quadratic growth law commenced on January 20, the
city of Wuhan was already under quarantine.
This suggests that the following thirty days of nearly perfectly
quadratic growth where solely the result of human interventions,
and therefore potentially highly unstable.
This is evidenced by the subsequent period of short exponential growth,
before the second period of quadratic growth commenced.
The present work has demonstrated that quadratic growth laws are
generally the result of partial confinement and that the maximum
possible infection levels, given the existing confinement measures,
have already been reached.
At present, the consequences of relaxing these measures cannot be easily
predicted, given that the almost universal lockdown in Europe and the
US has been unique in human history.

\subsection*{Acknowledgements}

I thank Bengt Gustafsson for inspiring discussions.
This work was supported in part through the Swedish Research Council,
grant 2019-04234.
I acknowledge the allocation of computing resources provided by the
Swedish National Allocations Committee at the Center for Parallel
Computers at the Royal Institute of Technology in Stockholm.

\vspace{5mm}
\noindent
The author claims no conflict of interests.


\newpage
\appendix
\section{Late quadratic growth}
\subsubsection*{Published separately in \cite{B20data}}

In \cite{B20}, it was shown that in a two-dimensional epidemic model, the
normalized averaged number of infections $\bra{I}$ grows quadratically
in time when the population is not strongly mixed and the local number
of infections $I$ is large compared to the number of susceptible ones $S$
that can still be infected.
This leads to what is known as peripheral growth, which is always
quadratic, but with a time constant that depends on the number of
hotspots that are surrounded by an individual front.

\begin{figure}[h]\begin{center}
\includegraphics[width=\columnwidth]{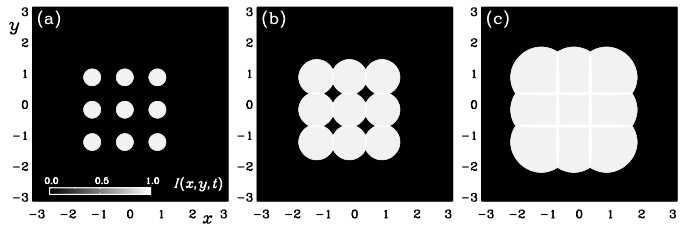}
\end{center}\caption[]{
Simulation with 9 hotspots that later merge and overlap.
The local distribution of $I(x,y,t)$ is shown in the $xy$ plane
for three values of $t$.
The length of the circumference determines the speed of growth.
When several hotspots merge, the circumference shortens and
the growth slows down.
}\label{psav_merge}\end{figure}

\begin{figure}[h]\begin{center}
\includegraphics[width=.99\columnwidth]{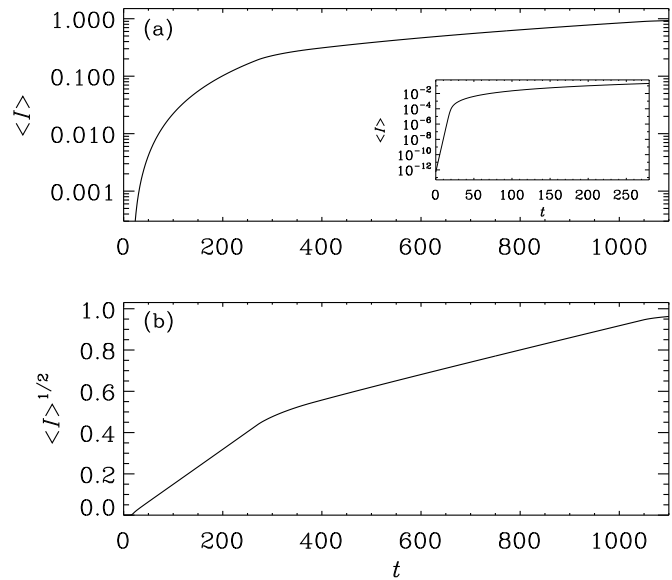}
\end{center}\caption[]{
Time series for simulation with 9 hotspots that later overlap.
Note that $N^{1/2}\propto\bra{I}^{1/2}$ grows linearly
with time $t$, which shows that $N\propto t^2$.
}\label{psqrt_merge}\end{figure}

It was shown that the growth is faster when there are more fronts,
and it was stated that the growth becomes slower when fronts merge
and several hotspots now connected by a common front.
The result of a simulation similar to the fiducial run of \cite{B20}
is shown in \Figs{psav_merge}{psqrt_merge}.
The value of $N$ is proportional to $\bra{I}$.
The simulations have been performed with the {\sc Pencil Code} \citep{PC}.

\begin{figure}[h]\begin{center}
\includegraphics[width=\columnwidth]{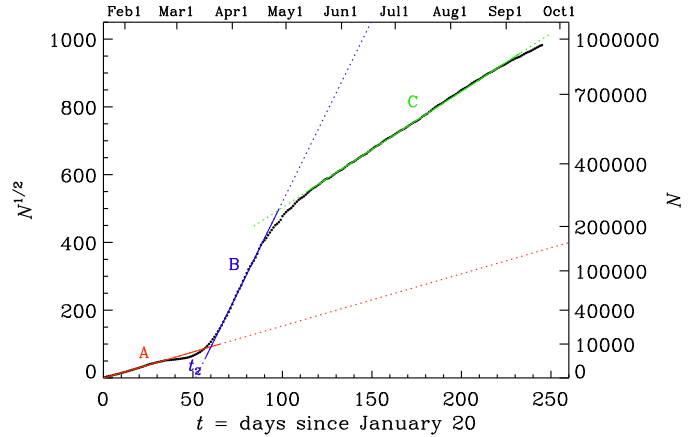}
\end{center}\caption[]{
Square root of $N$ (black dots) versus time since January~20, 2020.
The actual date is given on the upper axis and the actual values
of $N$ are given on the right-hand axis.
}\label{p3ext}\end{figure}

A decrease in the slope of $N^{1/2}$ versus $t$ is indeed seen since
May 2020; see section~C of \Fig{p3ext}, which is an updated version of
Figure~2 of \cite{B20}, where only data until April~20 were analyzed.
Here, the number of deaths, $N$, is taken from the worldometers
website\footnote{\url{http://www.worldometers.info/coronavirus/}}.
It is a more reliable proxy of the number of infections than the
reported number of infections.
During section~A, the epidemic was largely confined to China, but during
section~B many other places in the world got affected.

\end{document}